\def \preprint{Y}     % For preprint with epsf figures in the text.
\if \preprint Y \documentstyle[12pt,epsf]{article} \fi
\if \preprint N \documentstyle[12pt]{article} \fi
\newcommand{\beq}{\begin{equation}}
\newcommand{\eeq}{\end{equation}}

\begin{document}
\begin{titlepage}
\begin{flushright}
BI-TP 92/27, FSU-SCRI-92-101, HLRZ-92-40, TIFR/TH/92-42
\end{flushright}
\begin{flushright}
Bielefeld, July 22
\end{flushright}
\vspace{1.5cm}
\begin{center}
\large\bf{
A Lower Bound on $T_{SR}/{m_{\rm H}}$
in the O(4) Model on Anisotropic Lattices\
}\\
\vspace{1.5cm}
\large{R.V. Gavai,$^{1}$ U. M. Heller,$^{2}$ F. Karsch,$^{3,4}$\\
T. Neuhaus$^4$ and B. Plache$^4$
}
\end{center}
\vfill
\begin{center}
{\bf Abstract:}
\end{center}
Results of an investigation of the $O(4)$ spin model at finite temperature
using anisotropic lattices are presented. In both the large $N$
approximation and numerical simulations using the Wolff cluster algorithm
we find that the ratio of the symmetry restoration temperature $T_{\rm SR}$
to the Higgs mass $m_{\rm H}$ is independent of the anisotropy $\xi$. From
the numerical simulations we obtain a lower bound of $T_{\rm SR} / m_{\rm
H} \simeq 0.58 \pm 0.02$  at a value for the Higgs mass $m_{\rm H}a_s
\simeq 0.5$, which is lowered further by about $10\%$ at $m_{\rm H}a_s
\simeq 1$. Requiring certain timelike correlation functions to coincide
with their spacelike counterparts, quantum and scaling corrections to the
anisotropy are determined and are found to be small, i.e., the anisotropy is
found to be close to the ratio of spacelike and timelike lattice spacings.

\vfill
\footnotetext[1]{{\em
 Tata Institute of Fundamental Research, Homi Bhabha Road,
 Bombay 400005, India}}
\footnotetext[2]{{\em
 Supercomputer Computations Research Institute, The Florida
 State University, Tallahassee, FL 32306, USA}}
\footnotetext[3]{{\em
 HLRZ c/o Forschungszentrum J\"ulich, P. O. Box 1913,
 D-W-5170 J\"ulich, FRG}}
\footnotetext[4]{{\em
 Fakult\"at f\"ur Physik, Universit\"at Bielefeld,
 D-W-4800 Bielefeld, FRG}}
\end{titlepage}

\section{Introduction}
The fate of a spontaneously broken gauge theory at finite temperatures of
the order of the symmetry breaking scale has attracted attention for a
considerable period now.  Such investigations are of importance to the
physics of the very early universe. Two prime examples are the
inflationary universe and the generation of the baryon asymmetry. It has
been argued \cite{krs} that all the baryon asymmetry generated at the GUT
scale is washed out by non-perturbative effects near the electroweak phase
transition.  Whether any extra mechanism exists to create a fresh baryon
asymmetry \cite{ambjorn} near this phase transition remains unclear.
Although symmetry restoring phase transitions in spontaneously broken
gauge theories are crucial for these areas, our knowledge about them comes
chiefly from perturbation theory \cite{weilin} which can be expected to be
rather inadequate for dealing with the anticipated presence
of certain intrinsic non-perturbative effects
near such phase transitions \cite{linde}.

Motivated by the desire to learn more about the non-perturbative aspects of
the symmetry restoration transition, exploratory lattice investigations of
$SU(2)$ Higgs-gauge models and nonlinear $O(N)$ models at finite
temperature have been made \cite{Rly1,Rly2,jase,Nucu,schiestl}. These
models are expected to be trivial, giving rise to an upper bound on the
Higgs mass of about $650$ $GeV$ at values of the (lattice) Higgs mass close
to the lattice cutoff $m_{\rm H} a_s \simeq 1$. Correspondingly, at finite
temperature one expects a lower bound on the symmetry restoration
temperature $T_{\rm SR}$ in units of the inverse Higgs mass $m_{\rm
H}^{-1}$. For the standard model the gauge couplings at the weak symmetry
breaking scale and the Yukawa couplings, with the possible
exception of that of the top quark, are small.
Neglecting them as a first approximation,
one arrives at an $O(4)$ symmetric scalar model. As in the case
of the bound on the Higgs mass, one can then hope that the lower bound on
$T_{\rm SR}/m_{\rm H}$ can be obtained by studying the $O(4)$ model, rather
than the more complicated $SU(2)$ Fermion-Higgs model.

In this note we study the $O(4)$ model at finite temperature
using numerical simulation of the euclidian path integral on lattices with
anisotropic spacings in time (temperature) and spatial directions.
We also compare our results with analytic results obtained in the lowest
order large $N$ expansion. Anisotropic couplings allow, at least in
principle, a continuous tuning of the temperature, while the Higgs mass in
units of the lattice spacing can stay fixed at values $m_{\rm H}a_s \simeq
1$. Consequently a study of temperature effects of the theory at a
correlation length of the Higgs particle of order unity becomes feasible
without changing the lattice size or losing the resolution in the
temperature direction and the relevant information about the lower bound on
$T_{\rm SR}/m_{\rm H}$ can be extracted. In addition, anisotropic lattices
allow us to distinguish the finite temperature effects, which
in the euclidian formulation that we employ could be regarded
as a special type of finite size effects, from other finite
size effects since the finite temperature effects have to
be independent of the anisotropy in the scaling region.

The plan of this paper is as follows: In the next section we define the
model and give details of our methods to study it in the large $N$ limit
and, for $N = 4$, using numerical simulations. The procedure to obtain the
ratio $T_{\rm SR}/m_{\rm H}$ is described here. Section 3 is devoted to the
discussion of our results and conclusions are presented in the final
section. Some of our results have already been presented in a preliminary
form in \cite{rajiv}.
\newline

\section{The anisotropic O(N) model}

The anisotropic
$O(N)$ symmetric spin model on lattices with spatial
extension $N_s$ and temporal extension $N_t$
is defined by the action
\begin{equation}
S = - N \beta (\gamma \sum_x S_x \cdot S_{x + \hat 0} +
{1 \over \gamma} \sum_{x,j} S_x \cdot S_{x + \hat j} )
\label{eq:action1}
\end{equation}
or alternatively
\begin{equation}
S = - 2 \kappa (\gamma \sum_x S_x \cdot S_{x + \hat 0} +
{1 \over \gamma} \sum_{x,j} S_x \cdot S_{x + \hat j} ) .
\label{eq:action2}
\end{equation}
Here the spins $S_x$ are unit vectors in $O(N)$, $\gamma$ is the anisotropy
coupling and $\beta$ or $\kappa$ denote the hopping parameter.
Isotropic lattices are defined by
$\gamma = 1$. For the study of the large $N$ limit we take the first
form of the action, eq.~(\ref{eq:action1}),
keeping $\beta$ finite, while for the $O(4)$ model we
use the more conventional second form, eq.~(\ref{eq:action2}).

  Denoting the lattice spacing in spatial directions $a_s$ and in the
temporal direction $a_t$,
the anisotropy
parameter $\xi$ is the ratio of spacelike to timelike lattice spacings
\begin{equation}
\xi ={{a_s}\over{a_t}}.
\label{eq:xi}
\end{equation}

In the na\"\i ve continuum limit and for noninteracting theories $\xi =\gamma$.
However, quantum and scaling corrections can modify this relation
\cite{Frith}. For a given anisotropy coupling $\gamma$, $\xi$ can be
determined by the requirement that physics, e.g., the fall-off of
correlation functions, is the same in the time and spatial directions. As
the $O(4)$ model is weakly interacting we expect only a small renormalization
of $\xi$ with respect to the bare coupling $\gamma$. We also expect at the
critical point $\xi = \gamma$, as the renormalized coupling of the $O(4)$
model vanishes there. The anisotropy $\xi$ is easily calculable in the
large $N$ limit for the symmetric phase of the model. There we found that
the relevant contributions to $\xi$ are of order $O(a^2)$, i.e., a scaling
violation effect. The same conclusion can be inferred for the broken phase.
We also looked at contributions to $\xi$ in renormalized perturbation
theory of the $\lambda \phi^4$ model. Up to two loops we again found
$O(a^2)$ effects, in contrast to theories involving gauge fields where
quantum corrections of order $O(g^2)$ occur \cite{Frith}. We conjecture
that for our model the difference of $\xi$ from $\gamma$ is order $O(a^2)$
to all orders in perturbation theory.

On the anisotropic lattice the physical $3$-dimensional spatial volume and
the temperature are respectively given by $V_3=N_s^3a_s^3$ and $T=1/N_t
a_t=\xi/N_ta_s$. It is therefore  possible to vary $\xi$ and $N_t$
simultaneously at fixed ratio $\xi /N_t$, without changing the
temperature in units $a_s^{-1}$, or changing the spatial volume. This
amounts effectively to a change in resolution in the time direction: $a_t$
is changed while $N_t a_t$ is kept fixed. At least in the scaling region
physical results should then be independent of the anisotropy $\xi$. A
verification of this property will provide a valuable consistency check to
our analysis.

Earlier studies \cite{Rly2,jase} of the symmetry restoration phase
transition in the $O(4)$ symmetric spin model on isotropic lattices
revealed that it was only possible to determine the symmetry restoration
temperature $T_{\rm SR}$ for values of the Higgs mass which barely exceeded
a value of $m_{\it H} a_s \simeq 0.4$ on reasonable lattice sizes.
Furthermore, increasing the Higgs mass in units of the lattice spacing
$a_s$, one expects a logarithmically slow decrease of $T_{SR}/m_{\it H}$,
driving numerical simulations on isotropic lattices to larger temperatures
and smaller $N_t$ values, therefore loosing resolution in the time direction.
However choosing the anisotropy coupling $\gamma > 1$ it is possible to
explore the model at values $m_{\it H} a_s \simeq 1$ and at larger values
of the temperature $1/N_t a_t$ without giving up a reasonable
discretization in the time direction, i.e., in our case it was possible to
simulate the region $m_{\rm H}a_s \simeq 1$ on an $N_t=4$ lattice. In this
way it will be possible for the first time to explore regions of the theory
where the Higgs mass takes values of the order of the cutoff and a
numerical determination of the lower bound on $T_{SR}$ becomes feasible.

  In both the large $N$ calculation and the numerical simulations our
procedure to investigate finite temperature effects
consists of two steps. First we determine,
at given value of the anisotropy coupling $\gamma$, the critical
coupling on $N_s^3 \times N_t$ lattices.
Studying the large $N$ limit in leading order,
$\beta_c$ is obtained by solving numerically the
saddle point equation
\begin{equation}
\beta_c(N_t) =  {\gamma \over N_t N_s^3} \sum_p {}' {1 \over D(p) }
\label{eq:saddle}
\end{equation}
for $N_s \rightarrow \infty$,  where $D(p)$ is given by
\begin{equation}
D(p) = 4 \gamma^2 {\rm sin}^2 ({{1}\over{2}} p_0) +
4 \sum_j {\rm sin}^2 ({{1}\over{2}} p_j) ~~,~~
\label{eq:propa}
\end{equation}
with the momenta $p_\mu$ given by $p_\mu = 2 \pi n_\mu / N_\mu$, $n_\mu =
0, \dots , N_\mu - 1$, where $N_0 = N_t$ and $N_j = N_s$. The prime on
the sum in eq. (\ref{eq:saddle}) indicates that the zero mode $p = 0$
is being left
out. In Monte Carlo (MC) simulations the unique crossing point of the
Binder cumulant $g_{\rm R}= \langle M^4 \rangle/\langle M^2 \rangle^2$
for various
volumes $N_s^3$ and at given values of the anisotropy coupling $\gamma$
yields $\kappa_c(\infty, N_t)$. Here $M$ is the order
parameter, defined by
$M=\langle (M^{\alpha} M^{\alpha})^{0.5} \rangle $, where
$M^{\alpha}$ is given by
\begin{equation}
M^{\alpha}  = {{1}\over{N_s^3 N_t}}\sum_x  S_x^{\alpha}
\end{equation}
and $\alpha$ denotes the $O(N)$ index.
Alternatively, one may use the peak position of the susceptibility $\chi=
N_t N_s^3 (\langle M^2 \rangle - \langle M \rangle^2)$, to define
$\kappa_c(N_s,N_t)$. Using the critical exponents of the $O(4)$ model in
three dimensions, $\kappa_c(\infty,N_t)$ can then be obtained using
the finite size scaling theory.  We employed both  methods and checked
that they yield consistent results.

Secondly the Higgs mass and the renormalized field expectation value were
then determined at zero temperature at the coupling $\kappa_c(\infty,N_t)$
on $N_s^3 \times \gamma N_s$ lattices. For the determination of the
renormalized field expectation $v_{\rm R}$ in units of $a_t^{-1}$ we
proceed in case of our Monte Carlo simulation as follows: The dimensionless
quantity $v_{\rm R}a_t$ is given by an estimator for the field expectation
value $\Sigma$, which is properly normalized by its corresponding wave
function renormalization constant $Z$: $v_{\rm R}a_t=\Sigma / \sqrt{Z}$.
Note here that neither quantity $\Sigma$ nor $Z$ are fixed numbers in the
theory. It is possible to redefine $\Sigma$ and $Z$ by overall
multiplicative factors, such that the physical quantity $v_{\rm R}a_t$
stays fixed. In our case we chose the expectation value of the mean field
multiplied with a convenient factor $(\sqrt{2 \kappa} / \gamma) < M >$ as
an estimator for the field expectation value $\Sigma$. The corresponding
wave function renormalization constant can then be derived from the
behavior of the $O(4)$ symmetric zero momentum correlation function
\begin{equation}
G(n)={{2 \kappa} \over {4N_s^3 \gamma^2 }} {\sum_{\vec{x}}}
<S^\alpha_0 S^\alpha_{\vec{x},ne_t } >,
\label{eq:g}
\end{equation}
which is defined in the temporal direction of the lattice. Using chiral
perturbation theory one finds for large values of $n$ on a periodic
symmetric box, that $G(n)$ has the shape of a parabola. This is due to the
presence of massless Goldstone bosons in the theory: \begin{equation} G(n)=
Z {{3}\over{2V}} (n-{{N_t}\over{2}})^2 + {\rm const}. \end{equation}
Expressing the volume $V$ in units of $a_t$, $V=\xi^3N_s^3N_t$, the desired
wave function renormalization constant $Z$ can in principle be determined.
In our actual data analysis we have also considered the contribution of the
scalar particle  to eq.~(\ref{eq:g}); for a detailed description of the
procedure see \cite{o4pap}.

For the determination of the Higgs mass we project the scalar fields
$S^\alpha_x$ individually in each configuration onto the direction of the
mean field $M^{\alpha} /$ $ \mid M \mid$ and we obtain a field operator
which has a good overlap with the Higgs particle: \begin{equation}
S_{\sigma,x} = {{S^{\alpha}_x M^{\alpha}}\over{\mid M \mid}}.
\end{equation} The Higgs mass $m_{\rm H}a_t$ can then be extracted from the
exponential decay of the zero (spatial) momentum correlation functions of
the operator $S_{\sigma,x}$.

Introducing $O(N)$ invariant correlation functions defined on the main
axis of the lattice in time direction
\begin{equation}
C_t (n)= {{1}\over{N_s^3N_t}}\sum_x S^{\alpha}_x S^{\alpha}_{x+n e_t}
\end{equation}
and in space direction
\begin{equation}
C_s (n)= {{1}\over{N_s^3N_t}}\sum_x S^{\alpha}_x S^{\alpha}_{x+n e_s}
\end{equation}
we demand invariance with respect to an interchange of the spatial and
temporal directions. We match the correlation functions in temporal and
spatial directions at equal distance $n$, by scaling the temporal direction
by a factor $\xi$, which determines the anisotropy. As we shall see below,
the difference of $\xi$ from $\gamma$ was found to be rather small, being
of the order of at most $3 \%$ for all $\gamma$-values we studied.

\section{Results}

The numerical computations have been performed using the
nonlocal Wolff cluster algorithm. The employed statistics
were about $10^5$ sweeps for each simulated lattice size and set
of couplings. At finite temperature we simulated
$N_t$ and $\gamma$ values as given in Table 1.
In each case we performed simulations with $N_s=18$ and
$N_s=24$ at few values of the hopping parameter $\kappa$.
We employed the spectral density method in order
to determine the maximum of the susceptibility and the crossing point
of the Binder cumulant.
At zero temperature, with $\kappa=\kappa_c(\infty,N_t)$,
we performed simulations on $18^3 \times \gamma 18$ lattices
with $\gamma$ equal to the cited values.
\begin{table}[h]
\begin{center}
\begin{tabular}{||c|c|c|c||}                        \hline
 $N_t$ & $\gamma$ & $\kappa_c(\infty,N_t)$ & $\hat{\kappa}_c(\infty,N_t)$  \\
\hline
  6   &  1.~ & 0.3060(3)   & 0.314594     \\ \hline
  4   &  1.~ & 0.3103(3)   & 0.320871     \\ \hline
  6   &  1.5 & 0.3645(3)   & 0.377673     \\ \hline
  8   &  2.~ & 0.3912(3)   & 0.408333     \\ \hline
  3   &  1.5 & 0.3913(3)   & 0.415998     \\ \hline
  4   &  2.~ & 0.4171(3)   & 0.446373     \\ \hline
\end{tabular}
\end{center}
\caption[tab1]{\it Critical hopping parameters
at given $N_t$ and $\gamma$ for the $N_s \to \infty$ limit.
The third row  denotes our result from numerical simulations while
the last row ($\hat{\kappa}$) denotes results from the large $N$ expansion.}
\end{table}

Fig. 1 exhibits our results for both $g_{\rm R}$ and $\chi$ on $18^3 \times
6$ and $24^3 \times 6$ lattices for $\gamma =1.5$.   We used the spectral
density method to obtain the smooth curves shown from our data, shown by
crosses. Similar results have also been obtained for all other values of
$\gamma$ and $N_t$. In each case we obtained $\kappa_c(\infty,N_t)$ by
using both the crossing point of $g_{\rm R}$ and the finite size scaling of
the peak position of the susceptibility.  Both estimates were always found
to be consistent, although we preferred to use the former for determining
$m_{\rm H}$. Table 1 contains our results for $\kappa_c$ as a function of
$N_t$ and $\gamma$ from the numerical simulations, along with the
corresponding results from the large $N$ expansion. One finds a sizable but
less than $\sim 7\%$ difference between the two estimates, which is of the
same order as the discrepancy observed by comparing the zero temperature
critical hopping parameter from the large $N$ expansion with high precision
numerical simulations.

\if \preprint Y
\begin{figure}
\vspace{6in}
\includegraphics{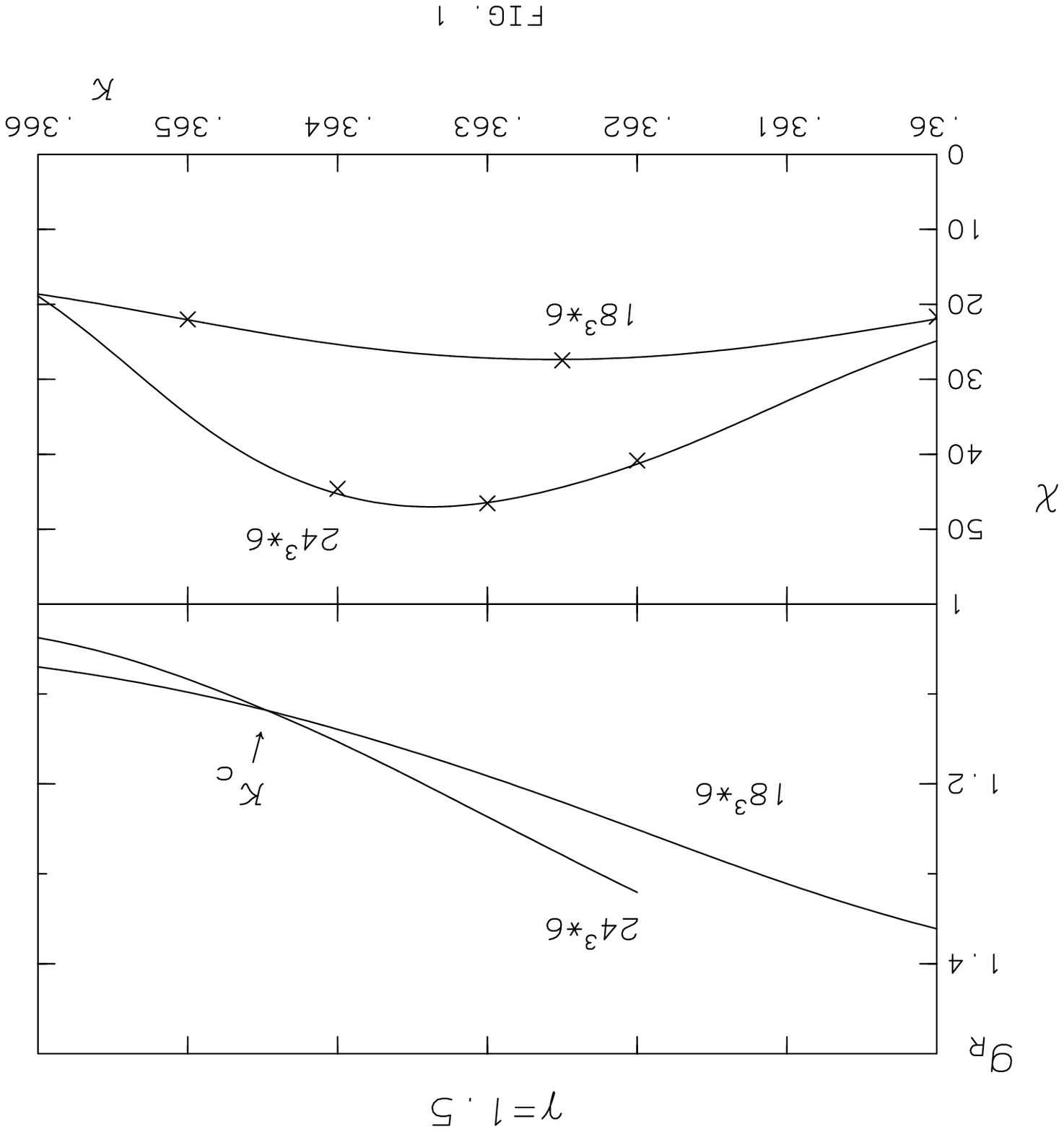}
\caption{Results for thermodynamic quantities at $\gamma=1.5$.}
\end{figure}
\fi

Fig. 2 compares the spacelike correlation function $C_s(n)$ on
an $18^3 \times 36$ lattice at ($\kappa$,$\gamma$) = (0.3912,2.0) with
the corresponding timelike correlation function $C_t(n/\xi)$ at scaled
distance $n/\xi$.  One sees that the two are in nice agreement
with each other. Table 2 contains, together with other
quantities, the measured anisotropy $\xi$. The deviations of $\xi$
from $\gamma$
are small, on the few percent level, which is in accord with the
expectations near a gaussian fixed point.
\begin{table}[h]
\begin{center}
\begin{tabular}{||c|c|c|c|c|c|c|c||}    \hline
$N_s$&$N_t$& $\gamma$&$\kappa$&$\xi$&$m_{\rm H}a_t$&$m_{\rm H}a_s$&
$T_{\rm SR}/m_{\rm H}$ \\
 & & & &$< M >$& Z & $v_{\rm R}a_t$ &
$T_{\rm SR}/v_{\rm R}$ \\ \hline
  18& 18&1.0&.3060&1.00(1)& 0.280(3)& 0.280(04)&.593(39) \\
 & & & & .1305(3) & 0.96(02) & .1040(14) & 1.60(10)  \\ \hline
  18& 18&1.0&.3103&1.00(2)& 0.428(3)& 0.428(07)&0.583(18) \\
 & & & & .2082(2) & 0.95(04) & .1682(37) & 1.485(64) \\ \hline
  18& 27&1.5&.3645&1.51(2)& 0.285(5)& 0.433(09)&0.583(30) \\
 & & & & .1913(1) & 0.96(03) & .1110(20) & 1.500(76) \\ \hline
  18& 36&2.0&.3912&2.05(2)& 0.212(1)& 0.436(06)&0.587(31) \\
 & & & & .1879(1) & 0.99(04) & .0832(15) & 1.501(91) \\ \hline
  18& 27&1.5&.3913&1.51(3)& 0.615(3)& 0.934(18)&0.541(07) \\
 & & & & .3838(1) & 0.97(05) & .2292(65) & 1.454(52) \\ \hline
  18& 36&2.0&.4171&2.05(5)& 0.457(5)& 0.937(25)&0.547(12) \\
 & & & & .3716(1) & 0.97(06) & .1719(61) & 1.453(66) \\ \hline
\end{tabular}
\end{center}
\caption[tab1]{\it Main results from the numerical
simulation of the finite temperature $O(4)$ model on anisotropic lattices.}
\end{table}
%\eject

\if \preprint Y
\begin{figure}
\vspace{6in}
\includegraphics{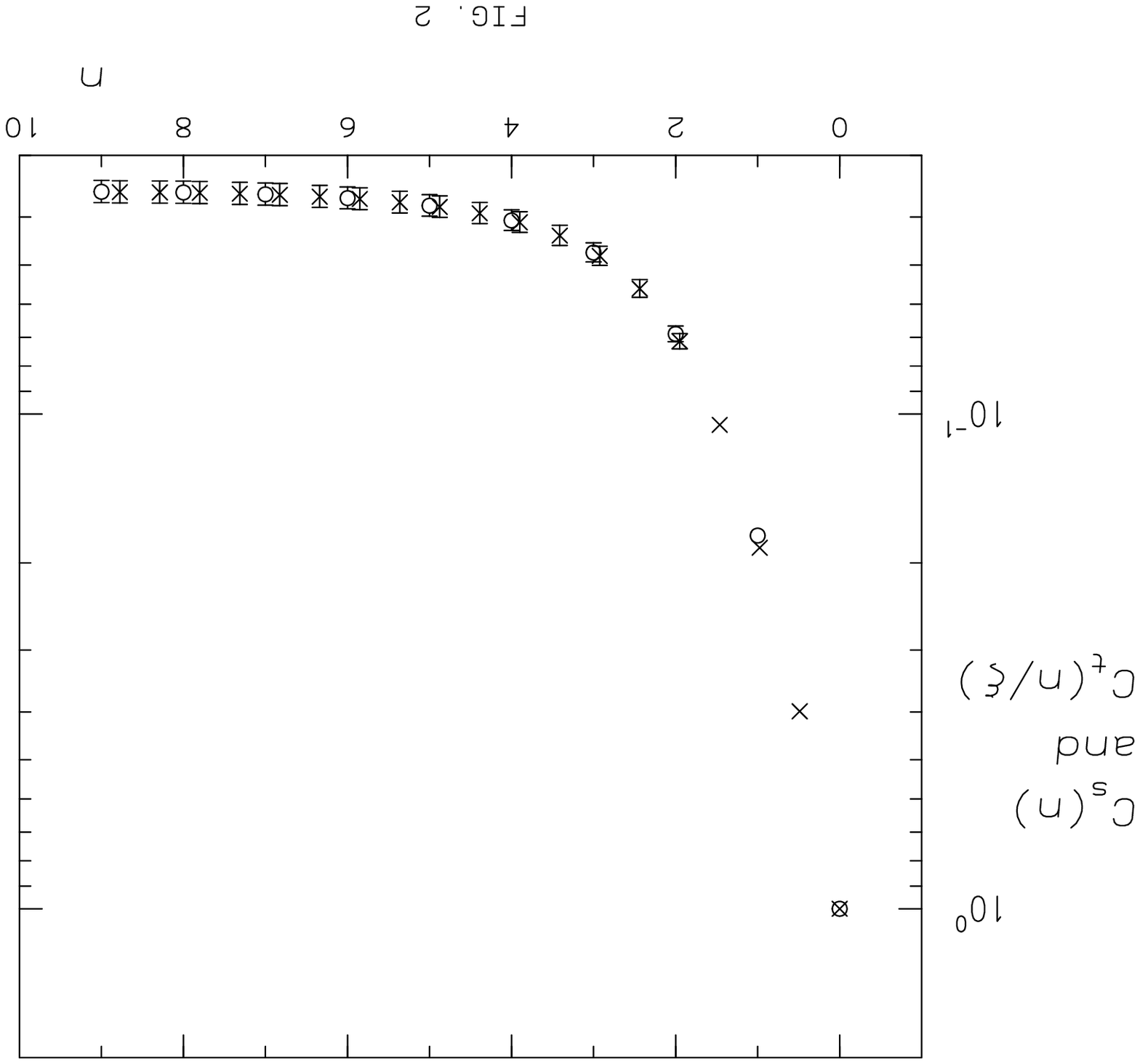}
\caption{$C_s(n)$ and $C_t(n/\xi)$ at ($\kappa,\gamma$)= (0.3912,2.0)
on a $18^3 \times 36$ lattice.}
\end{figure}
\fi

The Higgs mass $m_{\rm H}a_t$ was then obtained from an exponential fit to
the connected zero momentum correlation functions of the operator eq.~(9).
These values of the Higgs mass are listed together with $m_{\rm H}a_s$ in
Table 2. Using these results, the ratio $T_{\rm SR}/m_{\rm H} = 1/N_t
m_{\rm H}a_t$ shown in the table was obtained for various $\gamma$ and
$N_t$. As expected, we observe the $\xi$-independence of the ratio at fixed
values of $m_{\rm H}a_s$, demonstrating the internal consistency of our
finite temperature formulation of the theory on anisotropic lattices.
Considering fluctuations around the saddle point in the large $N$ limit,
one can obtain the Higgs mass $m_{\rm H}$ at $\beta_c(\infty,N_t)$ at given
$\gamma$. Fig. 3 shows these large $N$ results for $T_{\rm SR}/m_{\rm H}$.
They are also seen to be almost independent of the anisotropy $\xi$.

\if \preprint Y
\begin{figure}
\vspace{6in}
\includegraphics{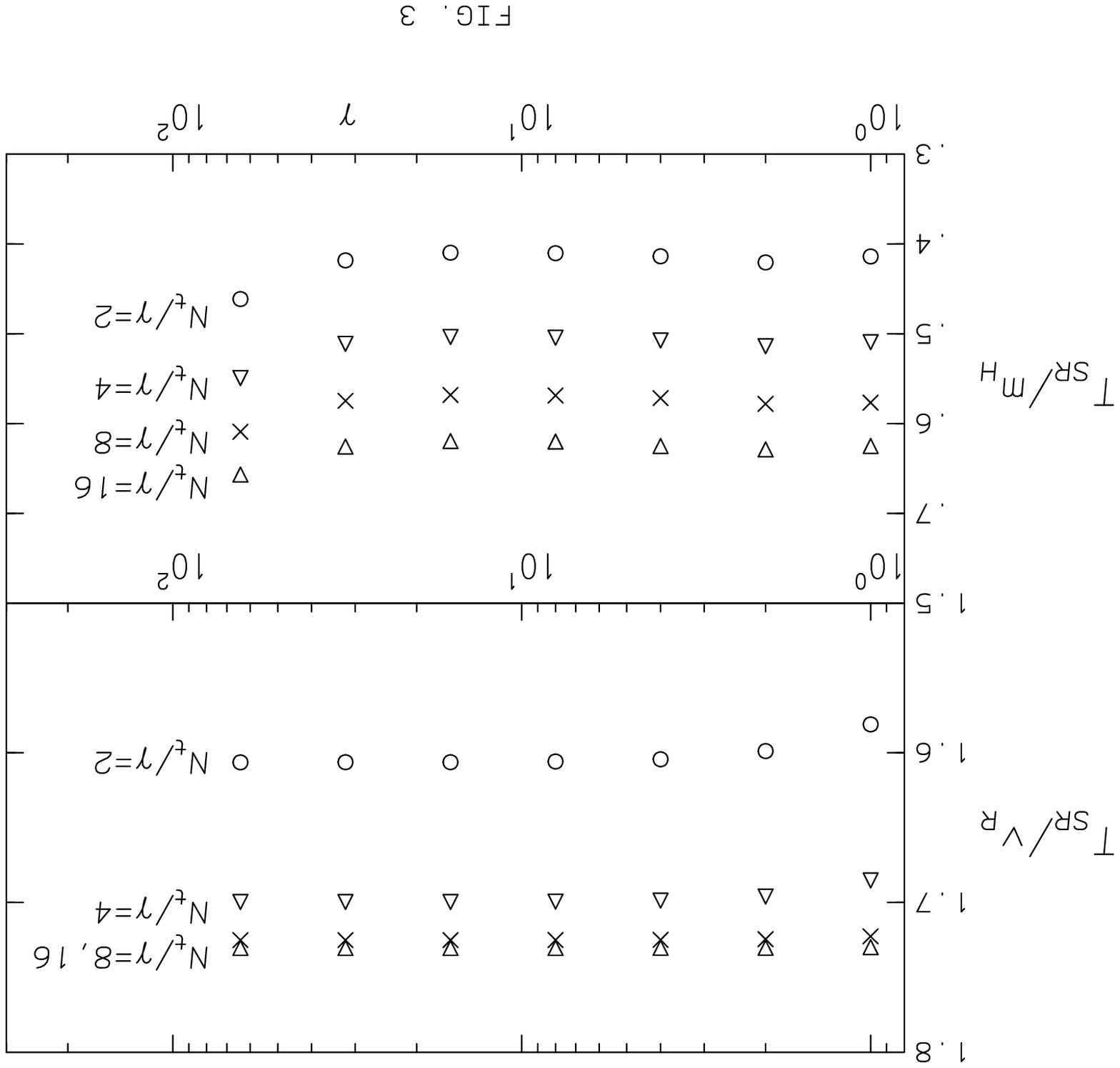}
\caption{Large $N$ results for $T_{SR}/v_{\rm R}$ and $T_{SR}/m_{\rm H}$ as
a function of the anisotropy parameter $\gamma$.}
\end{figure}
\fi

We have also collected in Table 2 our results for the expectation value of
the mean field $< M >$, the wave function renormalization constant $Z$,
$v_{\rm R}a_t$ and finally the ratio $T_{\rm SR}/v_{\rm R}$, which have
been determined from the Monte Carlo data by the methods described above.
Once again the $\xi$ independence of the ratio at fixed $m_{\rm H}a_s$ is
nicely born out. Large $N$ results for the ratio $T_{\rm SR}/v_{\rm R}$ are
also shown in Fig. 3. The renormalized vacuum expectation value of the
field in the large $N$ calculation is given by $v_{\rm R}^2 = N
(\beta_c(\xi L_t) - \beta_c(\infty))$ and we have set $N =
4$.$^1$ Again, the ratio $T_{\rm SR}/v_{\rm R}$ is almost independent of
the anisotropy $\xi$. The large $N$ results, shown in Fig.~3, agree quite
well with the numerical results at $N = 4$ of Table 2.

\footnotetext[1]{{The normalization of $v_R$ in \cite{rajiv} differs by a
factor $\sqrt{N}$ (= 2 for $N = 4$) from the one used here. This causes a
difference of a factor 2 in the scale of Fig.~2 there as compared to Fig.~3
in this paper.}}

A remark concerning the error determination for the quantities as cited in
Table 2 and a comment on further possible systematic errors may be
appropriate here. As can be noted, the ratios $T_{\rm SR}/m_{\rm H}$ and
$T_{\rm SR}/v_{\rm R}$ exhibit sizable errors, as compared to the
relatively small and purely statistical errors quoted for all the other
quantities. These errors are mainly caused by the uncertainty of the finite
temperature critical $\kappa$ values (Table 1), which in turn lead to
relatively large errors for the zero temperature values of $m_{\rm H}a_t$
and $v_{\rm R}a_t$ to be used in the ratios. Also we have to expect zero
temperature finite volume corrections to the quantities $m_{\rm H}a_t$ and
$v_{\rm R}a_t$ used to construct the ratios as quoted in Table 2. As we
anticipate the finite volume corrections to the quantities $T_{\rm
SR}/m_{\rm H}$ and $T_{\rm SR}/v_{\rm R}$, quoted in Table 2, to be
significantly smaller on our lattices than the errors induced by the
uncertainty of the critical points, we refrained from a detailed zero
temperature finite size scaling analysis. Future simulations yielding more
precise $\kappa_c$ values will have to incorporate them.

Our data for $T_{\rm SR}/m_{\rm H}$ as depicted in Table 2 decrease, as
expected, very slowly as the Higgs mass $m_{\rm H}a_s$ in units of $a_s$ is
increased. Thus, depending on the choice of value of the correlation length
up to which an effective theory can be defined, one obtains a lower bound
on the ratio $T_{\rm SR}/m_{\rm H}$. Just as in the case
of the determination of the
upper bound to the Higgs mass, it is expected that this lower bound
saturates for the theory under study, i.e., the $O(4)$ model at infinite
bare quartic coupling. From Table 2 we estimate this bound to be $0.58 \pm
0.02$ for a correlation length of $\sim  2$, which further decreases by
about $10\%$ for a value of $m_{\rm H}a_s \simeq 1$. Our data for $T_{\rm
SR}/v_{\rm R}$ show an approximate constant behavior as $m_{\rm H}a_s$ is
varied. The actual value is within the  errors consistent with the value
$\sqrt{2}$, which is the prediction of one-loop renormalized perturbation
theory, though the data show some tendency to lie slightly above the
perturbative value.

   It is interesting to compare our results for $T_{\rm SR}/m_{\rm H}$
with the one-loop result as obtained in renormalized perturbation
theory in the $O(4)$ model.
To this order the symmetry restoration temperature
is given by \cite{jase}
\begin{equation}
{T_{\rm SR} \over {m_{\rm H}}} =({{6}\over {g_{\rm R}}})^{1\over{2}},
\label{eq:pert}
\end{equation}
where $g_{\rm R}$ is the renormalized quartic coupling of the model.
Using previous high precision numerical determinations of
$g_{\rm R}$ \cite{japan} as an input we draw in Fig. 4 our numerical
results for $T_{\rm SR}/m_{\rm H}$ (crosses) together with the one-loop
prediction as indicated by the curve and by the triangles. Here
we observe sizable deviations when $m_{\rm H}a_s$ takes values
$\approx 1$, indicating that higher order corrections are
large at finite temperatures in a region of the model
where the scalar correlation length is close to $1$; see also \cite{kripf}.
Including also results from the large
$N$ expansion in Figure 4, one also notices sizable
deviations of the large $N$ results from our data, though the overall
trend is reproduced.

\if \preprint Y
\begin{figure}
\vspace{6in}
\includegraphics{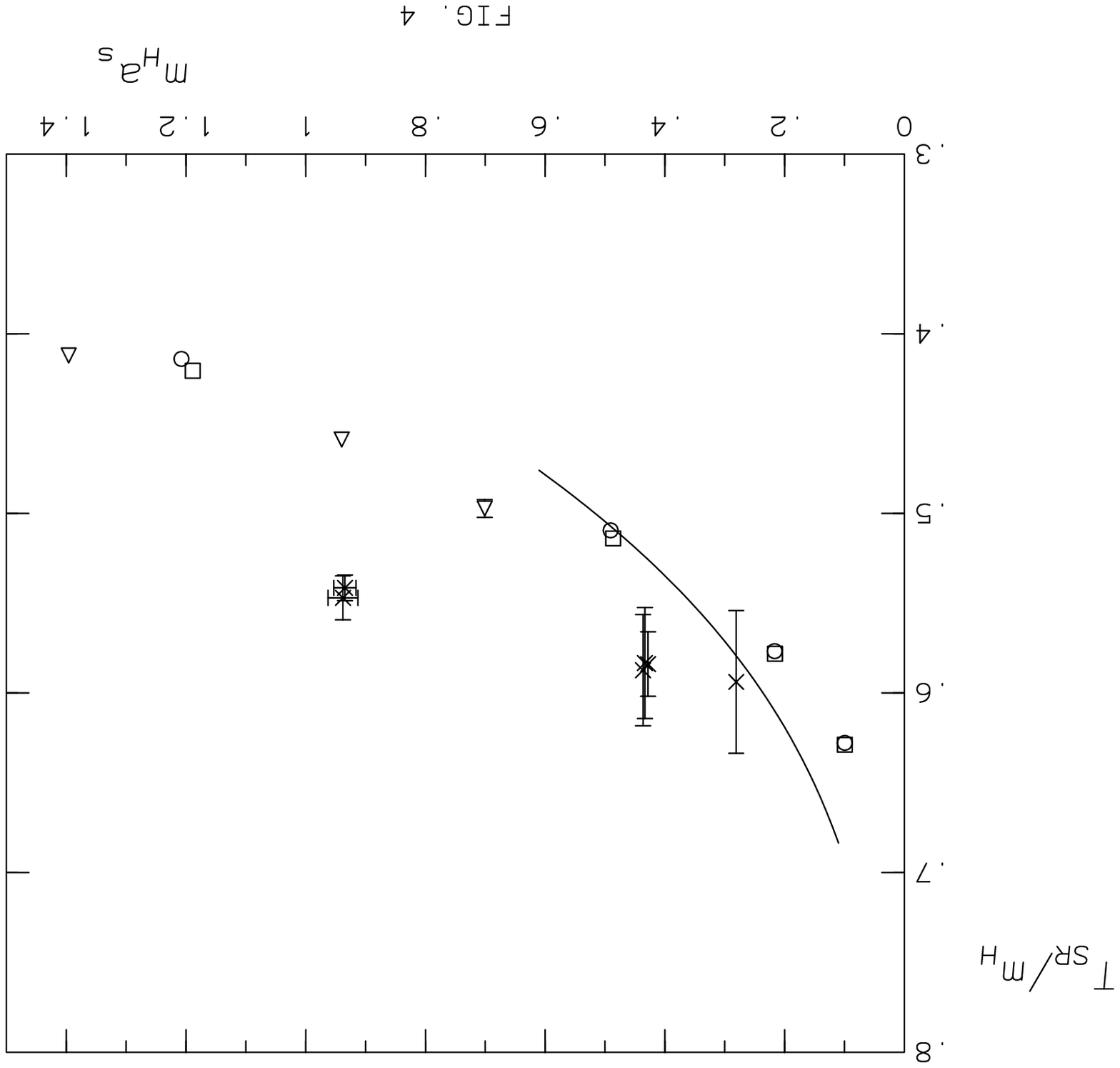}
\caption{$T_{\rm SR}/m_{\rm H}$ as a function of $m_{\rm H}a_s$. The
crosses denote our numerical results, circles and boxes come from the large
$N$ expansion at various values of $N_t$ with anisotropy parameter
$\gamma=1$ (circles) and $\gamma=2$ (boxes), while the curve and triangles
correspond to 1-loop renormalized perturbation theory, eq. (12).}
\end{figure}
\fi

\section{Conclusions}

Using anisotropic lattices we have studied the finite temperature behavior
of the $O(4)$ theory in regions of the parameter space where the
correlation length of the scalar particle is as low as $\approx 1$. Depending
on the maximal value of $m_{\rm H}a_s$ one is willing to admit for a
sensible definition of the effective theory, a lower bound on $T_{\rm
SR}/m_{\rm H}$ is derived. E.g., for a heavy Higgs particle which at a
value of the cutoff $m_{\rm H}a_s \simeq 0.5$ has a mass close to its
triviality bound of about $650~GeV$, we find $T_{\rm SR}=370~GeV$. This
value is close to the value predicted by renormalized perturbation theory
$T_{\rm SR}=\sqrt{2} v_{\rm weak}$ with $v_{\rm weak} \simeq 250~GeV$ and
consistent with our finding that the ratio $T_{\rm SR}/v_{\rm R}$ follows
the perturbative answer in the whole considered correlation length
interval. However, at correlation length $1$ we start finding large
deviations from one-loop perturbation theory for the quantity ${T_{\rm
SR}/m_{\rm H}}$. Qualitatively, the lowest order large $N$ expansion seems
to reproduce all the features of the Monte Carlo (MC) data well.
Even quantitatively the results are consistent with the na\"\i ve
expectation that they should be accurate to $O(1/N)$.
In the large $N$ expansion we
were able to explore $\xi$-independence of $T_{\rm SR}/m_{\rm H}$ and
$T_{\rm SR}/v_{\rm R}$ over larger ranges of $\xi$ and for more values of
$N_t$. This supports our belief that the early scaling evidence in the MC
data even for $N_t=3$ and 4 lattices is no fluke. But it would be
interesting to check this by simulating the theory at more $\xi$ and $N_t$
values.

{\bf Acknowledgments:} R.V.G. wishes to acknowledge the
financial support of the HLRZ J\"ulich where this work
was begun. He is also thankful
to its staff members for their
kind hospitality.
The work of U.M.H. was supported by the U.S. Department of Energy under
contract No. DE-FC05-85ER2500000.
He also wishes to acknowledge the hospitality
provided to him at University Bielefeld during the early stages
of this work.
The computations have been performed on the Cray at HLRZ.
\hfill\break

\vfill\eject

\vfill\eject

\if \preprint N
\section{Figure Captions}
{}.
\vskip1.0truecm
{\bf Figure 1:} Results for thermodynamic quantities at
$\gamma=1.5$.

\vskip1.0truecm
{\bf Figure 2:}
$C_s(n)$ and $C_t(n/\xi)$ at ($\kappa,\gamma$)= (0.3912,2.0)
on a $18^3 \times 36$ lattice.

\vskip1.0truecm
{\bf Figure 3:}
Large $N$ results for $T_{SR}/v_{\rm R}$ and $T_{SR}/m_{\rm H}$ as a function
of the anisotropy parameter $\gamma$.

\vskip1.0truecm
{\bf Figure 4:}
$T_{\rm SR}/m_{\rm H}$ as a function of $m_{\rm H}a_s$. The crosses
denote our numerical results, circles and boxes come from the large
$N$ expansion at various values of $N_t$ with anisotropy
parameter $\gamma=1$ (circles)
and $\gamma=2$ (boxes), while
the curve and triangles corresponds to 1-loop
renormalized perturbation theory, eq. (12).

\fi
\end{document}